\RequirePackage{fix-cm}
\documentclass[smallextended]{svjour3}
\smartqed
\usepackage{epsfig}
\usepackage{graphicx}
\usepackage[tbtags]{amsmath}
\usepackage{amssymb}

\begin{document}

\title{Equal-fidelity surface for a qubit state via equal-distance extended Bloch vectors and derivatives}

\subtitle{}

\author{Sang Min Lee \and Heonoh Kim \and Han Seb Moon}

\institute{
Sang Min Lee \at
Department of Physics, Pusan National University, Busan 609-735, Korea \\
\email{samini79@gmail.com}           
\and
Heonoh Kim \at
Department of Physics, Pusan National University, Busan 609-735, Korea
\and
Han Seb Moon \at
Department of Physics, Pusan National University, Busan 609-735, Korea
}

\date{Received: date / Accepted: date}

\maketitle

\begin{abstract}
We describe and characterize an equal-fidelity surface in Bloch space targeted for a qubit state  by means of equal-distance concept. The distance is generalized and defined as the Euclidean distance between extended Bloch vectors for arbitrary dimensional states. The distance is a genuine distance according to the definition and is related to other distances between quantum states and super-fidelity.
\end{abstract}

\section{Introduction}
Fidelity is the concept most frequently used to compare two quantum states in quantum information science, and related to various distances between the quantum states, even though it is not a metric. For examples, Bures distance is directly convertible to the fidelity~\cite{Bures,Hubner}, and fidelity gives upper and lower bounds of trace distance~\cite{Fuchs}. In the paper, we introduce a distance between two quantum states that is defined as the Euclidean distance between two extended Bloch vectors in $\mathbb{R}^{N^2}$ space. This distance was already proved to be a metric, namely, the ``modified root infidelity,'' in a previous work~\cite{Miszczak}. However, we give a simple geometrical definition of the distance. From the distance between two qubit states ($N=dim(\mathcal{H})=2$), we can directly calculate the fidelity. However, for the cases of $N>2$, the derived quantity is an upper bound of the fidelity (super-fidelity)~\cite{Miszczak}.

The beginning of this research is a curiosity about geometry of equal-fidelity states in Bloch space targeted for a qubit state, because two well-known examples show completely different features. For a pure state and the maximally mixed state, equal-fidelity states are represented by an orthogonal plane to the Bloch vector of the target state and a sphere of which center is on the target state, respectively. We try to explain the two different aspects into a unified mechanism (equal-distance of extended Bloch vectors) through a mediate example, a non-maximally mixed state.

The paper is structured as follows. Section~\ref{BN} briefly introduces the fidelity and the generalized Bloch vector for $N$ $\ge$ 2. In Sec.~\ref{EF}, we first show two aforementioned examples of equal-fidelity surfaces, then describe schematically how to obtain an equal-fidelity surface for a general qubit state and describe its properties. In Sec.~\ref{DEB}, we discuss the distance between the extended Bloch vectors for $N\ge2$ and the relationships with fidelity. Finally, a summary is given in Sec.~\ref{SUM}.

\section{Basic notation} \label{BN}

\subsection{Fidelity}
Fidelity is the concept most frequently used to compare two quantum states in quantum information science because it has legitimate properties~\cite{Miszczak,Jozsa}:
\begin{itemize}
  \item Bounds: $0\leq F(\rho_1,\rho_2) \leq 1$,
  \begin{itemize}
  \item $F=1$ iff $\rho_1=\rho_2$,
  \item $F=0$ iff $supp(\rho_1) \bot supp(\rho_2)$.
  \end{itemize}
  \item Symmetry: $F(\rho_1,\rho_2)=F(\rho_2,\rho_1)$.
  \item Unitary invariance: $F(\rho_1,\rho_2)=F(U\rho_1U^\dagger,U\rho_2 U^\dagger)$.
\end{itemize}
The fidelity is defined as
\begin{align}
F(\rho_1 , \rho_2) &\equiv \left ( Tr \left[ \sqrt{\sqrt{\rho_1}\rho_2\sqrt{\rho_1}}  \right ]   \right)^2 \label{FD1}\\
&= Tr[\rho_1 \rho_2] + 2 \sum_{i<j} \chi_i \chi_j, \label{FD2}
\end{align}
where $\{\chi_i\}$ are the eigenvalues of the matrix $\sqrt{ \sqrt{\rho_1} \rho_2 \sqrt{\rho_1} }$~\cite{Miszczak}. The physical meaning and derivation of the fidelity can be found in previous works on the transition probability~\cite{Uhlmann1976} and purification of mixed states~\cite{Jozsa}.

\subsection{Generalized Bloch vector} \label{GBV}
The Bloch vector is a very common representation of a density matrix $\rho$ for a qubit state~\cite{Nielsen}. The Bloch vector is defined as $\vec
 {\lambda} =(\lambda_x,\lambda_y,\lambda_z)$ in $\mathbb{R} ^3$, where $\lambda_i=Tr[\rho \hat \sigma_i]/2$, and satisfies $|\vec {\lambda}|\leq 1/2$. Then the density matrix is expressed\footnote{The maximum length of the Bloch vector (for pure states) can be modified by adopting a constant $\alpha$ as $\rho = I/2 + \alpha \sum_{i=1} ^3 \lambda_i \hat \sigma_i$. Although the usual notation is $\alpha=1/2$ so that $\lambda_i=Tr[\rho \hat \sigma_i]$ and $\left|\vec \lambda \right| \leq 1$, we set $\alpha=1$ in this paper for a consistent argument later.}  as $\rho =  I/2 +  \sum_{i=1} ^3 \lambda_i \hat \sigma_i$, where $\{\hat \sigma_i\}$ are the Pauli operators.

For an $N$-dimensional Hilbert space, the Bloch vector is generalized via generators $\{\hat \lambda _1, \cdots, \hat \lambda _{N^2-1}  \}$ of $SU(N)$. In this paper, the generalized Bloch vector~\cite{Miszczak,Kimura,Byrd} is defined as $\vec {\lambda}^{(N)} =(\lambda_1, \cdots, \lambda_{N^2-1})$ in $\mathbb{R} ^{N^2-1}$, where $\lambda_i=Tr[\rho \hat \lambda_i]/2$, and the density matrix is expressed as $\rho = I/N + \sum_{i=1} ^{N^2-1} \lambda_i \hat \lambda_i$. The length of the generalized Bloch vector is bounded as $\left|\vec {\lambda} ^{(N)}\right|\leq \sqrt{\frac{N-1}{2N}}$ (equality for only pure states). The generators of $SU(N)$ are defined as
\begin{gather} \label{G}
\{ \hat \lambda _i \}_{i=1} ^{N^2-1} =\{\hat u _{jk}, \hat v _{jk}, \hat w _m \},  \\
\hat u _{jk}  = |j\rangle\langle k| + |k\rangle\langle j|, \notag \\
\hat v _{jk}  = -i|j\rangle\langle k| + i|k\rangle\langle j|, \notag \\
\hat w _{m}  = \sqrt{\frac{2}{m(m+1)}} \left( \sum_{j=1} ^{m} | j\rangle\langle j| - m|m+1\rangle\langle m+1| \right), \notag
\end{gather}
where $1 \leq j \leq k \leq N$ and $1 \leq m \leq N-1$. They satisfy
\begin{gather}
\hat \lambda _i  = \hat \lambda _i ^\dagger, ~~~~
Tr [\hat \lambda _i]  = 0, ~~~~
Tr [\hat \lambda _i \hat \lambda _j]  = 2 \delta _{ij}. \label{GP}
\end{gather}
The operators $\hat u _{jk}$, $\hat v _{jk}$, and $\hat w _m$ are generalized Pauli operators of $\hat \sigma_x$, $\hat \sigma_y$, and $\hat \sigma_z$, respectively.

\section{Equal fidelity surface for a qubit state in Bloch space} \label{EF}
\begin{figure}[t]
\centering
\includegraphics*[scale=0.4]{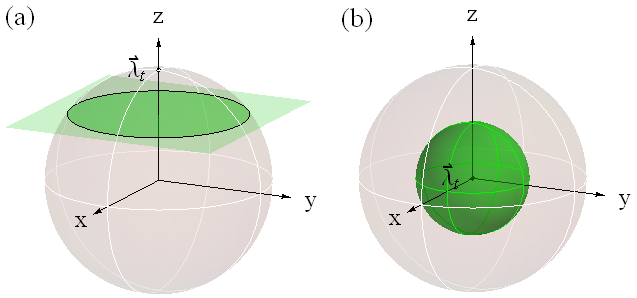}
\caption{Representations of equal-fidelity states in Bloch space, when the target state is (a) a pure state: case {\bf{A}}, $\left|\vec \lambda_t \right| =1/2$, and (b) the maximally mixed state: case {\bf{B}}, $\left|\vec \lambda_t\right| =0$.}
\label{FIG1}
\end{figure}

In general, density matrices of which the Bloch vectors are located at the same distance from a target vector $\vec \lambda _t$ have different fidelities with the target state $\rho_t$. For a simple example, ({\bf{A}}) equal-fidelity states for a pure target state are represented by an orthogonal plane of the target vector in Bloch space, as shown in Fig.~\ref{FIG1} (a). The reason is as follows. When one of the states to be compared is pure, the fidelity is written in a simple form as
\begin{align} \label{FL1}
F(\rho_t ,\rho) & = Tr\left[ \rho_t \rho \right]
 = \frac{1}{2} + 2~\vec \lambda_t  \cdot \vec \lambda
\end{align}
using Eq.~(\ref{FD1}) and Eq.~(\ref{GP}). Therefore, a set of $ \{ \vec \lambda  \}$ on an orthogonal plane for  $\vec \lambda_t$ has the same fidelity with the target state. On the other hand, ({\bf{B}}) for the maximally mixed state ($I/2$), a set of equal-fidelity states are located at the same distance from the origin of the Bloch space, as shown in Fig.~\ref{FIG1} (b). Since the fidelity and the maximally mixed state are unitary invariant,
\begin{gather} \label{FL0}
F(I/2,\rho)=F(I/2,U\rho U^\dagger)
\end{gather}
is satisfied. A unitary operation corresponds to a rotation operation in Bloch space, so Eq.~(\ref{FL0}) means that a set of $\{ \vec \lambda\}$, which are equivalent under rotation (have the same length), has the same fidelity with the target state $I/2$ (the origin of the Bloch space)\footnote{When the target state is the maximally mixed state for $N=2$ (qubit), the same fidelity states are equivalent to the same purity states (the same Bloch vector lengths). However, for $N>2$, having the same purity states is a sufficient condition for having the same fidelity states for the target $I/N$. For example, $\rho_a=diag(0.735, 0.1325, 0.1325)$ and $\rho_b=diag(0.04, 0.48, 0.48)$ have different purities but the same fidelity with the target $\rho_t=(1/3, 1/3, 1/3)$.}. These two extreme examples have completely different features (flat plane for $\left|\vec \lambda_t \right| = 1/2$ and sphere for $\left|\vec \lambda_t \right| = 0$).

To investigate general cases, $0 \leq \left|\vec \lambda_t \right| \leq 1/2$, we adopt modified forms of the fidelity and Bloch vector. The fidelity between two qubits can be represented by their Bloch vectors~\cite{Hubner,Jozsa} as
\begin{gather} \label{FD2D}
F(\rho_1,\rho_2)=\frac{1}{2}+ 2 \vec \lambda_1 \cdot \vec \lambda_2  + 2 \sqrt{(1/2)^2-\left|\vec \lambda_1 \right|^2} \sqrt{(1/2)^2-\left|\vec \lambda_2 \right|^2} .
\end{gather}
If we extend the Bloch vector into  $\vec L =  \left(\lambda_x,\lambda_y,\lambda_z,\sqrt{(1/2)^2-\lambda_x ^2 -\lambda_y ^2 - \lambda_z ^2} \right)$~\cite{Hubner,Miszczak,Uhlmann1992,Sommers}, then Eq.~(\ref{FD2D}) is modified concisely as
\begin{gather} \label{FD3}
F(\rho_1,\rho_2)=\frac{1}{2} + 2 \vec L_1 \cdot \vec L_2.
\end{gather}
The extended Bloch vector $\vec L$ is on a hyperhemisphere of $S^3$, i.e., $| \vec L | = 1/2$ and $0 \leq L_4$, namely, an ``Uhlmann hemisphere.'' Equation~(\ref{FD3}) shows that a set of equal-fidelity states is represented by a hyperplane in $\mathbb{R} ^4$, similar to the case of a pure target state in Eq.~(\ref{FL1}) and Fig.~\ref{FIG1} (a). However, $ \{ \vec L \}$ are restricted on the hyperhemisphere of $S^3$, so the solution is given by the intersection between the hyperplane and the hyperhemisphere.

In other words, the fidelity can be described by the distance between the two extended Bloch vectors. Since the Euclidean distance of two vectors is represented as $| \vec L _1 - \vec L _2  |^2  =  1/2  - 2  \vec L _1 \cdot \vec L _2$, the fidelity is rewritten as
\begin{gather}
F(\rho_1,\rho_2)  = 1 - | \vec L _1 - \vec L _2  |^2  \label{FFD}
\end{gather}
using Eq.~(\ref{FD3}). The above relation shows that fidelity between two qubit states is represented by the Euclidean distance between extended Bloch vectors of the states\footnote{This distance differs from the Bures distance.}.

\begin{figure}[b]
\centering
\includegraphics*[scale=0.4]{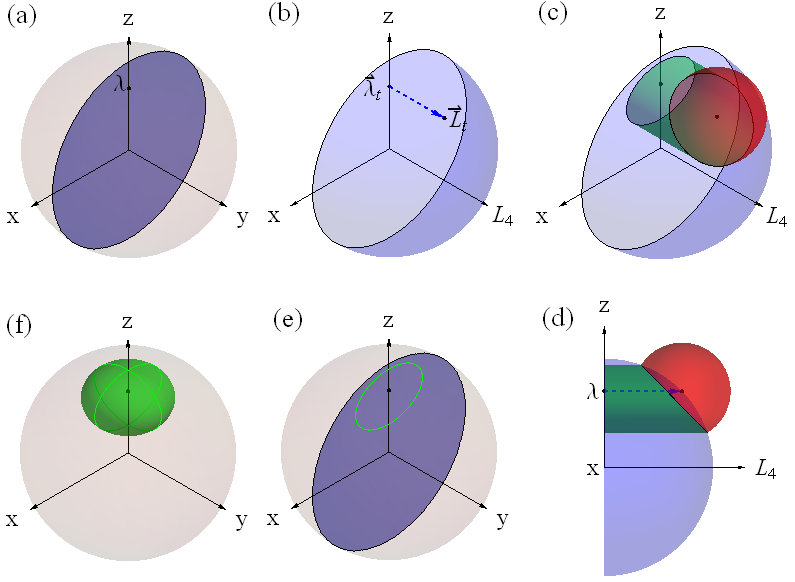}
\caption{Schematic diagram illustrating how to obtain an equal-fidelity surface in Bloch space. (a) $xz$ plane and target state $\vec \lambda_t=(0,0,\lambda \ge 0)$ in Bloch space. (b) Plane and target vector $\vec \lambda_t$ are represented by the hemisphere and the vector $\vec L_t$  in the extended Bloch space. (c, d) The set of equal-distance extended Bloch vectors $\{ \vec L \}_{ed}$ from the target vector $\vec L_t$ is represented by the intersection of two (blue and red) spheres in $\mathbb{R}^4$. (e) Set of vectors $\{ \vec L \}_{ed}$ is represented by a green ellipse on $xz$ plane in Bloch space. (f) The entire set of equal-fidelity states is represented by an ellipsoid in Bloch space using $z$ axis rotational symmetry.}
\label{rBeB}
\end{figure}

Figure~{\ref{rBeB}} shows schematic diagrams illustrating how to represent equal-fidelity states in Bloch space from equal-distance extended Bloch vectors. First, we assume that the target state is on the $+z$ axis as $\vec \lambda_t=(0,0,\lambda \ge 0)$, without loss of generality (by a unitary transformation). Then we consider the $xz$ plane in Bloch space, which is represented by the blue disk in Fig.~\ref{rBeB} (a). When we ignore $y$ axis because of $\lambda_y$ = 0, the disk is converted to a hemisphere in the extended Bloch space in Fig.~\ref{rBeB} (b).  The extended Bloch vector of the target $\vec L _t$ is projected on the hemisphere from $\vec \lambda_t$ along the $L_4$ direction, as shown in Fig.~\ref{rBeB} (b). A set of vectors $\{\vec L\}_{ed}$, which are located at the same distance from the target $\vec L_t$, is represented by the intersection of the blue hemisphere and a red sphere in Fig.~\ref{rBeB} (c) and (d). As shown in Fig.~\ref{rBeB} (e), the equal-distance extended Bloch vectors $\{\vec L\}_{ed}$ for $\vec L_t$ are represented by the green ellipse on the $xz$ plane of the Bloch space through the reverse projection. This argument can be applied to an arbitrary disk plane in the Bloch space that contains the origin $(0,0,0)$ and the target $(0,0,\lambda)$: $z$ axis rotation symmetry. Thus, the total set of equal-fidelity states generally has the form of an ellipsoid in Bloch space, as shown in Fig.~\ref{rBeB} (f). This schematic explanation in Fig.~\ref{rBeB} clearly shows the reason for the different features of the two examples: ({\bf{A}}) $\lambda = 1/2$ and ({\bf{B}}) $\lambda = 0$. The projected solutions in Fig.~\ref{rBeB} (e) for cases ({\bf{A}}) and ({\bf{B}}) are a straight line and a circle, respectively.

\begin{figure}[t]
\centering
\includegraphics*[scale=0.4]{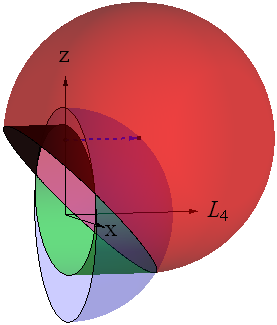}
\caption{When $D^2(\vec L _t, \vec L) \ge 1/2-\lambda$, there are spurious solutions. Since the extended Bloch vectors are restricted by $L_4>0$, the projection in black ($z > \frac{2F-1}{4 \lambda}$) represents spurious solutions for $L_4<0$.}
\label{nsol}
\end{figure}

For the target vector $\vec \lambda_t=(0,0,\lambda)$, where $0\leq\lambda\leq1/2$, the explicit expression of equal-fidelity states in Bloch space is
\begin{gather}
\frac{x^2+y^2}{F(1-F)}+\frac{\left(z-(2F-1)\lambda \right)^2}{F(1-F)(1-4\lambda^2)}=1,
\label{sol}
\end{gather}
where $z\leq \frac{2F-1}{4\lambda}$. The vectors on the ellipsoid where $z > \frac{2F-1}{4\lambda}$ are spurious solutions for the cases of $L_4<0$, as represented by the black projection in Fig.~\ref{nsol}.

\begin{figure}[t]
\centering
\begin{tabular}{ccc}
(a)&(b)&(c)\\
\includegraphics*[scale=0.175]{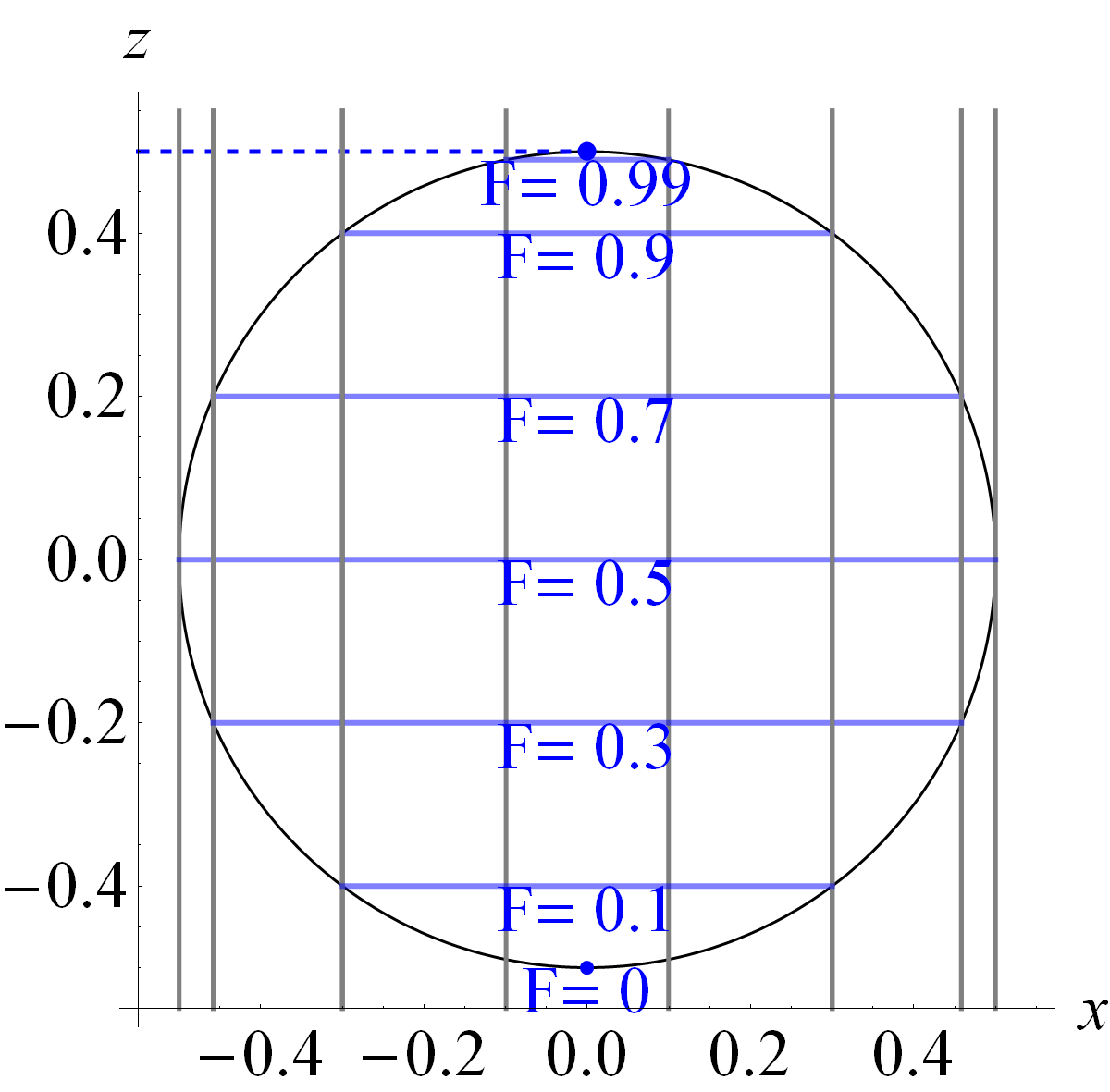}&
\includegraphics*[scale=0.175]{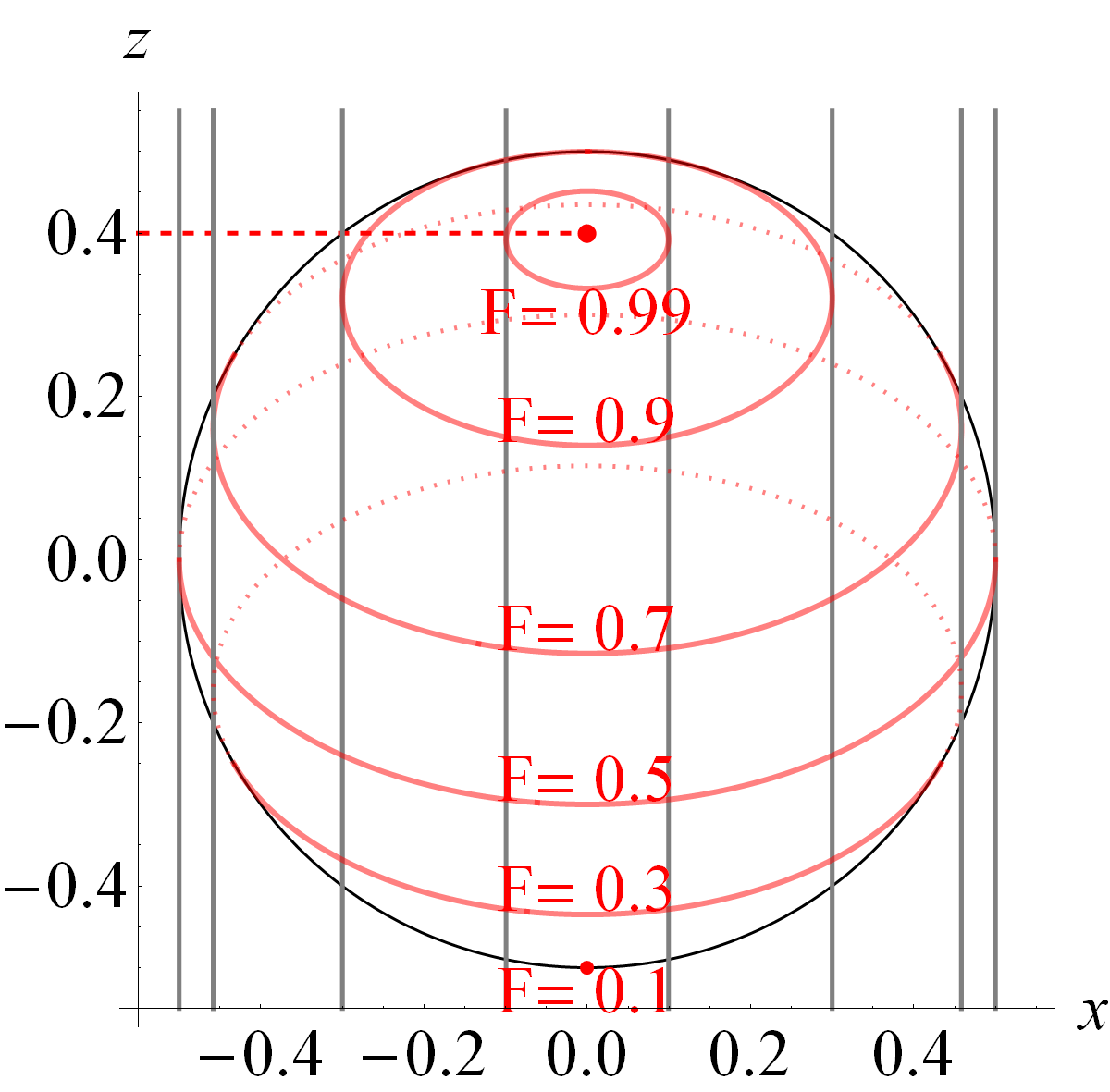}&
\includegraphics*[scale=0.175]{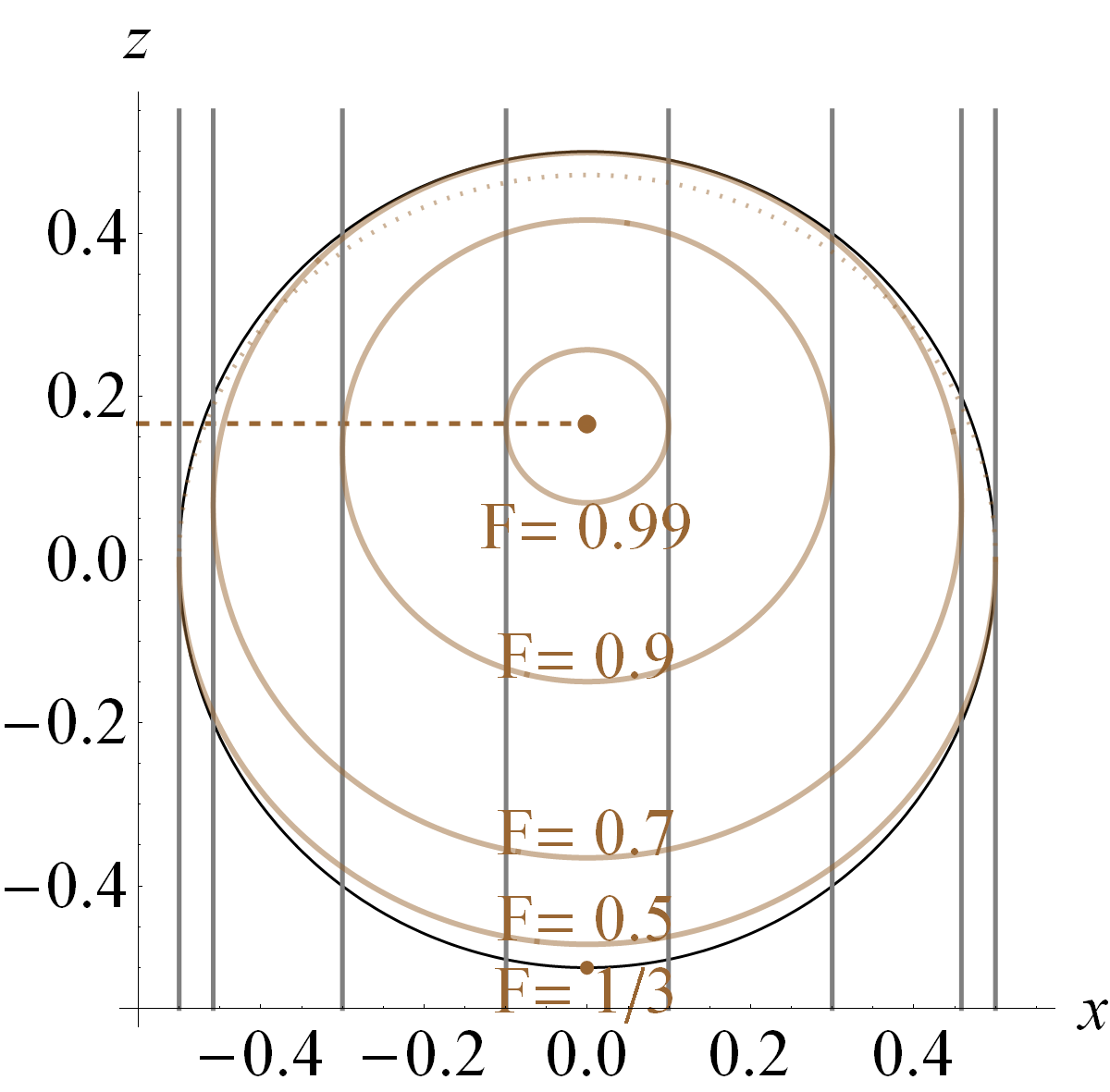}
\end{tabular}
\caption{Equal-fidelity states on $xz$ plane in Bloch space for (a) $\lambda=1/2$, (b) $\lambda=2/5$, and (c) $\lambda=1/6$. Large dots are target states; small dots are minimum fidelity states. Dotted lines in (b) and (c) represent spurious solutions.}
\label{xz}
\end{figure}

The oblate ellipsoid solution (major axis: $xy$ plane, minor axis: $z$ axis) in Eq.~(\ref{sol}) has two properties. The length of the semimajor axis is a function of the fidelity as $\sqrt{F(1-F)}$, and the ratio between the major and minor axes is fixed as $\sqrt{1-4\lambda^2}$. As shown in Fig.~\ref{rBeB} (c) and (e), the solution (green ellipse on $xz$ plane) for equal-fidelity states is projected from a circle on a tilted plane (not shown in Fig.~\ref{rBeB}) in the extended Bloch space. Since the length of the major axis is not affected by the tilt angle of the plane, it is determined by the distance in $\mathbb{R}^4$ or the fidelity. The ratio of the major and minor axes is a function of the tilt angle $\theta$ (angle between $\vec L_t$ and the $L_4$ axis), so it is given by $cos\theta=\sqrt{1-4\lambda^2}$. In Fig.~\ref{xz}, we show examples ($\lambda=$1/2, 2/5, and 1/6) of equal-fidelity states on the $xz$ plane of the Bloch space. They clearly show that the major axis is fixed as the fidelity and the eccentricities of the ellipses are fixed as $\lambda$, the length of the target Bloch vector.

In general, we think that two states are very close if their fidelity is 0.99. However, as shown in Fig.~\ref{xz} (a), the angle between two pure states for $F=0.99$ is about $11.5^\circ$. If we consider a simple experiment using a polarization qubit system, it corresponds to about a $5^\circ$ operational error of a half-wave plate. The experimental errors of wave plates are typically less than $1^\circ$ and could be further reduced via motorized rotation mounts. Recently, we experimentally demonstrated operational error-insensitive approximate universal-NOT gates in a polarization qubit system~\cite{Lee}. In the experiment, we measured the error-insensitivity of the gate via the fidelity deviation when the target state is an ideally flipped pure state $|\psi _\bot\rangle$ (comparable to $\lambda$ = 1/2) rather than an ideal output state $\rho'=\frac{1}{3}\sum_i \hat \sigma_i |\psi\rangle\langle\psi|\hat\sigma_i$ (comparable to $\lambda$ = 1/6) of the approximate UNOT gate, because the fidelities between the erroneous outputs and $\rho'$ are very close to unity. In other words, when the target is $\rho'$ ($\lambda$ = 1/6) in Fig.~\ref{xz} (c), the fidelity deviations of the erroneous outputs are very small, since most erroneous outputs are located inside the surface of $F=0.99$. Thus, the fidelity deviation can be changed by the target state, even though the distribution of states in Bloch space remains.

\section{Distance between extended Bloch vectors for arbitrary $N$} \label{DEB}
The extended Bloch vector $\vec L$ for a qubit state ($N=2$) is defined as $\left| \vec L \right| = 1/2$  by adding a fourth component. Similarly, an extended Block vector for $N\ge2$ can be defined using the generalized Bloch vector and one additional term $L_{N^2}$, as follows:
\begin{align}
\vec L & \equiv \left(\lambda_1,\cdots,\lambda_{N^2-1}, L_{N^2} \right), \label{DGEBV1} \\
L_{N^2} &\equiv \sqrt{\frac{N-1}{2N}-\left|\vec \lambda \right|^2} \notag \\
&=\sqrt{(1-Tr[\rho^2])/2}, \label{L}
\end{align}
so $\vec L$ is on a hyperhemisphere of $S^{N^2-1}$ with a radius of $\sqrt{\frac{N-1}{2N}}$. Using the Euclidean distance between the extended Bloch vectors for an arbitrary dimension, we define a distance $D_L$ between two density matrices as,
\begin{gather}
D_L(\rho _1, \rho _2) =  \left|  \vec L (\rho_1) - \vec L (\rho_2)  \right|. \label{DD}
\end{gather}
We redefine $F'$ between two density matrices\footnote{If we define the Bloch vector as $\rho=\frac{I}{N} + \frac{1}{2} \sum _i
\lambda_i \hat \lambda_i $,  where $\lambda_i = Tr[\rho \hat \lambda_i]$, then $L_{N^2} =  \sqrt{\frac{2(N-1)}{N}-\left| \vec \lambda \right|^2}$, and $F'(\rho_1,\rho_2)$ is defined as $1-D^2(\vec L_1,\vec L_2)/4$.}
similar with Eq.~(\ref{FFD}) as,
\begin{align}
F'(\rho_1,\rho_2) & \equiv 1 - D^2_L (\rho_1,\rho_2) \notag \\
&  = \frac{1}{N}+2\left(\vec \lambda_1 \cdot \vec \lambda_2 + \sqrt{\frac{N-1}{2N}- \left| \vec \lambda_1 \right|^2} \sqrt{\frac{N-1}{2N}-\left|\vec \lambda_2 \right|^2}\right) \notag \\
& = Tr[\rho_1 \rho_2] + \sqrt{\left(1-Tr[\rho_1^2] \right)\left(1-Tr[\rho_2^2] \right)},  \label{DF'} \\
D_L(\rho_1, \rho_2) & = \sqrt{1-F'(\rho_1,\rho_2)}. \label{DF''}
\end{align}
$F'$ has the following properties. It is unitary invariant, since the overlap of two density matrices and the purity of a density matrix are unitary invariant. When at least one of the density matrices $\rho_i$ is a pure state, $F'$ is the same as the fidelity between the two states, because the square root term in Eq.~(\ref{DF'}) is zero, and the fidelity is reduced to $Tr[\rho_1\rho_2]$ in that case. However, in general, it is an upper bound of the fidelity, namely, the ``super-fidelity''~\cite{Miszczak}.

As the definition itself (the Euclidean distance between two vectors in $\mathbb{R}^{N^2}$), $D_L$ satisfies the general properties of a distance:
\begin{itemize}
  \item Non-negativity: $d(x_1,x_2)\ge0$, $d=0$ iff $x_1=x_2$.
  \item Symmetry: $d(x_1,x_2)=d(x_2,x_1)$.
  \item Triangle inequality: $d(x_1,x_2)+d(x_2,x_3) \ge d(x_1,x_3)$.
\end{itemize}
In previous works~\cite{Miszczak,Gilchrist}, $D_L$ is represented by the ``modified root infidelity,'' $C'(\rho_1,\rho_2)$, and is proved to be a genuine distance in a different way.

We should note that the distance $D_L$ in Eq.~(\ref{DF''}) differs from the Bures distance, which is defined as $D_B^2=2-2\sqrt{F(\rho_1,\rho_2)}$~\cite{Bures,Hubner} even for $N=2$. However, if we assume that the fidelity is close to unity, as for $F=1-\delta$ ($\delta \ll 1$) and cases where $F=F'$, then $D_L^2$ and the Taylor-approximated $D_B^2$ are the same as $\delta$.

Now, we consider the inner distance  $\tilde{D}_L$ between two extended Bloch vectors. Since $\vec L_i$ are limited to the hyperhemisphere of $S^{N^2-1}$, the inner distance is defined using the length of the vectors and the angle $\theta$ between two vectors as
\begin{align}
\tilde{D}_L\left(\rho_1,\rho_2\right) & \equiv \left| \vec L_i \right| \theta \notag \\
& = \sqrt{\frac{N-1}{2N}} ~ cos^{-1} \left( \frac{2N}{N-1}\vec L_1 \cdot \vec L_2 \right) \notag \\
& = \sqrt{\frac{N-1}{2N}} ~ cos^{-1} \left( \frac{N F' -1 }{N-1}  \right), \\
&= \frac{1}{2} cos^{-1} \left( 4 \vec L_1 \cdot \vec L_2 \right) ~~~~~~ \textrm{for} ~ N=2 \notag \\
&= \frac{1}{2} cos^{-1} \left( 2 F(\rho_1,\rho_2) -1 \right)  \notag \\
&= cos^{-1} \sqrt{F(\rho_1,\rho_2)}, \label{BL}
\end{align}
using Eqs.~(\ref{FD3}) and (\ref{DF'}). For $N=2$, the inner distance between two extended Bloch vectors $\tilde D_L$ is the same as the Bures length~\cite{Nielsen,Sommers,Uhlmann1995}, as shown in Eq.~(\ref{BL}), and $\tilde D_L ^2$ becomes $\delta$ when $F=1-\delta$ $(\delta \ll 1)$.

\section{Summary}\label{SUM}
In this paper, we obtained the general expression for the equal-fidelity surfaces of a qubit state in Bloch space via the concept of equal distances of the extended Bloch vectors and explained the properties of the equal-fidelity surfaces. We generalized the extended Bloch vectors and their distance for an arbitrary-dimensional Hilbert space. The distance is a genuine distance according to the definition itself. From the distance between the extended Bloch vectors, we define $F'$ for the two density matrices. In general, $F'$ is an upper bound of the fidelity, although $F'$ is reduced to the fidelity in restricted cases, i.e., $N=2$ or at least one of the states is pure. We also show that $D_L$ is related to the Bures distance and Bures length. The distance $D_L$ is not a new distance between quantum states, but we introduce a definition in a new and intuitive way. We expect and hope that our research will facilitate further work on basic studies of the fidelity and quantum distances in quantum information science.

\begin{acknowledgements}
This work was supported by the National Research Foundation of Korea (NRF) grant funded by the Korea government (MSIP) (No. 2015R1A2A1A05001819 and No. 2014R1A1A2054719)\end{acknowledgements}

\end{document}